\documentclass{article}
\usepackage[utf8]{inputenc}
\usepackage[T1]{fontenc}

\usepackage{graphicx}
\usepackage{color}
\usepackage{amsmath}
\usepackage{amssymb}
\usepackage{hyperref}
\usepackage{comment}
\usepackage{authblk}

\newcommand{\hoarefol}{\textit{Hoare-fol}}

\title{The \hoarefol{} Tool}
\author{Maxime Folschette}
\affil{Univ. Lille, CNRS, Centrale Lille, UMR 9189 -- CRIStAL -- Centre de Recherche en Informatique Signal et Automatique de Lille, F-59000 Lille, France}

\begin{document}

\maketitle

\begin{abstract}
This document presents the tool named “Application of Hoare Logic and Dijkstra's Weakest Proposition Calculus to Biological Regulatory Networks Using Path Programs with Branching First-Order Logic Operators” or \hoarefol{} for short.
This tool consists in an implementation of the theoretical work developed in \cite{Bernot2019Genetically} and
contains the following features:
(1) computation of the weakest precondition of a Hoare triple,
(2) simplification of this weakest precondition using De Morgan laws and partial knowledge on the initial state, and
(3) translation into Answer Set Programming to allow a solving of all compatible solutions.
\end{abstract}

\section{Introduction}

\subsection{Biological Regulatory Networks}

Algebraic models \cite{Kauffman1969Metabolic, Thomas1973Boolean, DeJong2002Modeling} are noteworthy in the field of systems biology for their ease of use.
Indeed, contrary to other formalisms such as ordinary differential equation-based models, their complexity remains very low as they do not require to compute an analytical solution.
Furthermore, they require much less parameters to function, meaning that they are of great help when too many system parameters are unknown, while still yielding results on the modeled system's behavior.

However, less parameters does not mean no parameters.
As a consequence, finding one or several acceptable sets of parameters can still be a challenge, especially if the model is big and that this task cannot be tackled by hand.

\begin{figure}
  \centering
  \includegraphics[width=.5\linewidth]{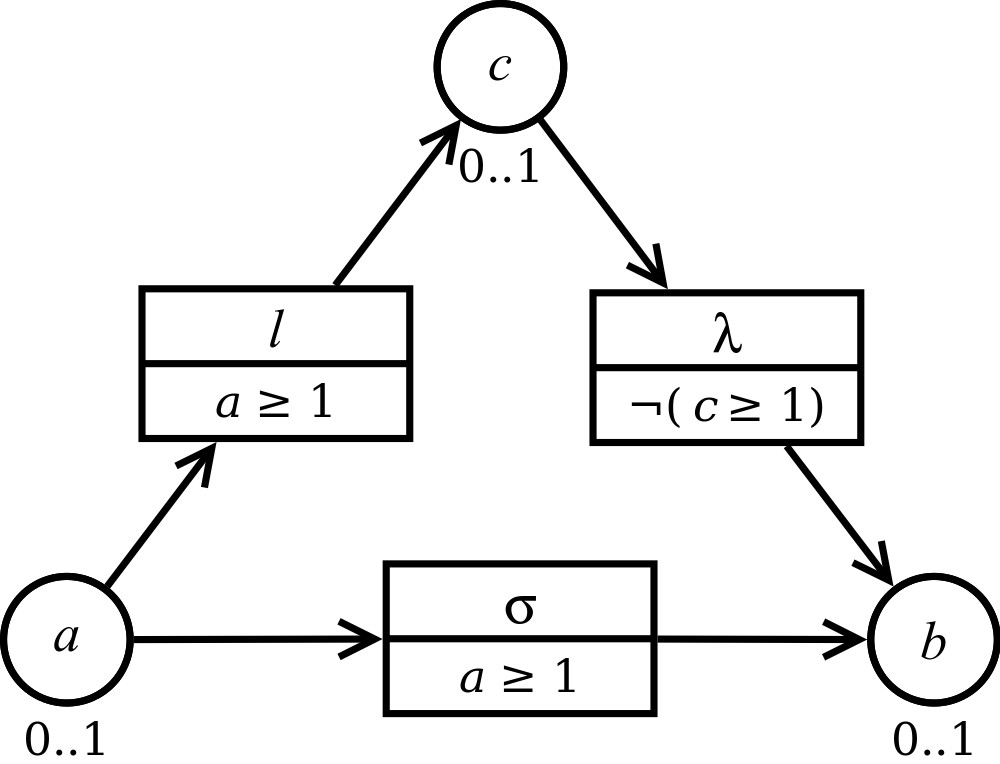}
  \caption{\label{fig:toy-thomas}%
    Toy example of \cite{Bernot2019Genetically} representing an incoherent feedforward loop.}
\end{figure}

The focus of this work is on \emph{Thomas' formalism}, which is typically used to represent Biological Regulatory Networks (BRNs) consisting in interacting components such as genes, proteins, external influences...
Formally, it takes the form of a graph in which nodes model components with discrete expression levels and edges model the interactions between these components.
More precisely, a specific extension of this formalism is considered, featuring hyperarcs labeled with logic formulas and called \emph{multiplexes}, that are useful to reduce the number of parameters \cite{Khalis2009Gene}.
An example of such a graph is given in Figure~\ref{fig:toy-thomas}.

\subsection{Parameters in Thomas' Formalism}

Yet, mutiplexes are not sufficient to specify some aspects of the dynamics:
\begin{itemize}
  \item how interactions play together when several of them point to the same node (that is, which logical gate is used),
  \item in the case of multi-valued models, how “strong” an interaction is (for instance, does it attract the component to level 1 or to level 2),
  \item in the case where it is not specified in the graph, whether an interaction “pulls” a component up (activation) or down (inhibition).
\end{itemize}

To represent this information, parameters for Thomas' formalism were proposed in \cite{Snoussi1989Qualitative} and are now considered as part of the formalism. They are often denoted with the letter $k$ and are uniquely characterized by a couple $(v, \omega)$ where $v$ is the variable it refers to and $\omega$ is the set of its active predecessors.
In other words, when the active predecessors of $v$ (that is, having an effective influence on $v$) are exactly the set $\omega$, then this variable $v$ is attracted towards the expression level $k_{v, \omega}$.

A set of parameters covering the whole model is called a \emph{parametrization}, and allows to compute, in each possible state, the global “focal point” towards which the model is dynamically attracted.
A parametrization is equivalent to a complete set of logic gates (and, or, etc.)\ between interactions towards the same node, and can also be equivalently translated into an activation function that takes the current state as input, and outputs the set of possible next states.
The parametrization representation is preferred here because it is central to Thomas' formalism and has been tailored to this kind of representaiton.

\subsection{Context of the \hoarefol{} tool}

The present work relies on the work developed in \cite{Bernot2019Genetically} which aims at providing a way to filter out unwanted parametrizations for a given model, based on known possible dynamical behaviors.
This work relies on the classical Hoare logic \cite{Hoare1969Axiomatic} by adapting it to Thomas' formalism: imperative programs become dynamical path programs (that represent a possible dynamical behavior of the model), and pre- and postconditions on the program's variables become conditions on the initial and final states and on the parametrization.
It also relies on Dijkstra's weakest preconditon calculus \cite{Dijkstra1978Guarded} to compute such information.

The rest of this docuemnt describes an implementation of this work under the form of the tool \hoarefol.
This implementation is written in OCaml and allows an export of the results to Clingo's Answer Set Programming~(ASP) \cite{Gebser2016Theory, Baral1994Logic} in order to enumerate all solutions.

This tool is a follow-up to the work started in \cite{Folschette2011Application}, where two unsuccessful approaches were taken:
\begin{itemize}
  \item using Coq to formally prove the new Hoare logic developed,
  \item using OCaml with functions to encode formulas (pre- and postconditions).
\end{itemize}
Both methods were not suited for weakest precondition calculus and manipulation, thus giving impractical or partial results.
This document, however, proposes a working proof-of-concept of such an implementation.

\section{Implementation}

The general idea of this implementation is to represent all conditions (pre- and postdconditions, and conditions of multiplexes) and dynamical path programs as symbolic trees in OCaml, by defining constructs for each type of node.
This representation allows to easily manipulate them in order to perform precondition calculus, simplification, translation to ASP, and so on.

Note however that in the current version of the implementation, all information (model, conditions, processings) must be provided as OCaml definitions in the main program\footnote{Which is of course neither developer-friendly nor a good practise, but this tool only intends to be a proof-of-concept.}.

\subsection{Model, Program and Formula Definitions}

First, the Thomas model takes the form of two lists representing the component nodes (\texttt{vars}) and multiplex nodes (\texttt{mults}) with all related information (predecessors, conditions, etc.).
Each element of these lists are couples where the first element is a string giving the name of the component (variable or multiplex) and the second is the information attached.
In the case of \texttt{vars}, the second element is also a couple where the first element is the upper bound of the variable (integer) and the second is the list of precedessor multiplex names (list of strings).
Regarding \texttt{mults}, the second element is the \emph{multiplex formula}, which follows the grammar given in Figure~\ref{fig:grammar-mult} and which itself contains information about its predecessors (variables or multiplexes).
For instance, the model of Figure~\ref{fig:toy-thomas} taken from \cite{Bernot2019Genetically} is represented in OCaml as the listing of Figure~\ref{fig:toy-ocaml}.

\begin{figure}
  \centering
  \noindent
  $\begin{array}{r@{\ }ll}
    \varphi :==
        & \texttt{MPropConst}(b)
          & \text{Constant proposition: $b$ is either True or False} \\
      | & \texttt{MPropUn}(n, \varphi)
          & \text{Unary proposition: $n$ is the negation} \\
      | & \texttt{MPropBin}(o, \varphi, \varphi)
          & \text{Binary proposition: $o$ is a connective ($\land$, $\lor$, ...)} \\
      | & \texttt{MRel}(c, \psi, \psi)
          & \text{Comparison: $c$ is a comparator ($=$, $>$, $\geq$, ...)} \\
      | & \texttt{MAtom}(v, i)
          & \text{Atom on a variable: means $(v \geq i)$} \\
      | & \texttt{MMult}(m)
          & \text{Atom on a multiplex: recalls the formula of $m$}
  \vspace*{.5em} \\
    \psi :==
        & \texttt{MExprBin}(o, \psi, \psi)
          & \text{Arithmetic operation: $o$ is an operator ($+$ or $-$)} \\
      | & \texttt{MExprConst}(i)
          & \text{Constant: $i$ is an integer}
  \end{array}$
  \caption{\label{fig:grammar-mult}%
    OCaml grammar for multiplex formulas ($\varphi$) and multiplex arithmetic expressions ($\psi$).}
\end{figure}

\begin{figure}
  \texttt{
    \begin{tabular}{r@{}l}
      let vars = [
        & ("a", (1, [])) ; \\
        & ("b", (1, ["lambda" ; "sigma"])) ; \\
        & ("c", (1, ["l"]))] ;; \\
      let mults = [
        & ("l", MAtom("a", 1)) ; \\
        & ("lambda", MPropUn(Neg, MAtom("c", 1))) ; \\
        & ("sigma", MAtom("a", 1))] ;;
   \end{tabular}
  }
  \caption{\label{fig:toy-ocaml}%
    OCaml representation of the toy example of \cite{Bernot2019Genetically}.}
\end{figure}

Then, general-purpose \emph{formulas} can be defined with another grammar defined in Figure~\ref{fig:grammar-formula} which is close to the multiplex grammar.
Such formulas will be used to define postconditions in order to perform weakest precondition calculus, but they could also be used to describe other kinds of conditions.
As a matter of fact, they are also used to express conditions and invariants in \texttt{If} and \texttt{While} control flow instructions.
For instance, the postcondition $(a = 1 \land b = 0)$ can be expressed with:
$$\texttt{\begin{tabular}{r@{}l}
  let my\_post = Pr&opBin(And, \\
    & Rel(Eq, ExprVar "a", ExprConst 1),\\
    & Rel(Eq, ExprVar "b", ExprConst 0)) ;;
\end{tabular}}$$

\begin{figure}
  \centering
  \noindent
  $\begin{array}{r@{\ }ll}
    \Phi :==
        & \texttt{PropConst}(b)
          & \text{Constant proposition: $b$ is either True or False} \\
      | & \texttt{PropUn}(n, \Phi)
          & \text{Unary proposition: $n$ is the negation} \\
      | & \texttt{PropBin}(c, \Phi, \Phi)
          & \text{Binary proposition: $o$ is a connective ($\land$, $\lor$, ...)} \\
      | & \texttt{Rel}(r, \Psi, \Psi)
          & \text{Comparison: $c$ is a comparator ($=$, $>$, $\geq$, ...)} \\
      | & \texttt{FreshState}(\Phi)
          & \text{Formula on a fresh set of variables}
  \vspace*{.5em} \\
    \Psi :==
        & \texttt{ExprBin}(o, \Psi, \Psi)
          & \text{Arithmetic operation: $o$ is an operator ($+$ or $-$)} \\
      | & \texttt{ExprVar}(v)
          & \text{Valuation of variable: the value of variable $v$} \\
      | & \texttt{ExprParam}(v, \omega)
          & \text{Valuation of parameter: the value of parameter $k_{v, \omega}$} \\
      | & \texttt{ExprConst}(i)
          & \text{Constant: $i$ is an integer}
  \end{array}$
  \caption{\label{fig:grammar-formula}%
    OCaml grammar for formulas ($\Phi$) and arithmetic expressions ($\Psi$) to be used in general-purpose conditions (preconditions, postconditions and control flow instructions).}
\end{figure}

Finally, an imperative \emph{path program} can be defined with the grammar given in Figure~\ref{fig:grammar-prog} which copies classical imperative program instructions (assignments) and control flow (\texttt{If}, \texttt{While}) but also adds descriptions for possible ($\exists$) and mandatory ($\forall$) dynamical branchings.
As an example, the path program $(b+; c+; b-)$ can be expressed with:
$$\texttt{let my\_prog = Seq(Seq(Incr "b", Incr "c"), Decr "b") ;;}$$

\begin{figure}
  \centering
  \noindent
  $\begin{array}{r@{\ }ll}
    \Pi :==
        & \texttt{Skip}
          & \text{Does nothing} \\
      | & \texttt{Set}(v, i)
          & \text{Assignment: $v \leftarrow i$} \\
      | & \texttt{Incr}(v)
          & \text{Increment: $v+$, i.e., $v \leftarrow v + 1$} \\
      | & \texttt{Decr}(v)
          & \text{Decrement: $v-$, i.e., $v \leftarrow v - 1$} \\
      | & \texttt{Seq}(\Pi, \Pi)
          & \text{Instructions sequence} \\
      | & \texttt{If}(\Phi, \Pi, \Pi)
          & \text{If-then-else conditional} \\
      | & \texttt{While}(\Phi, \Phi, \Pi)
          & \text{While loop: requires a condition and a loop invariant} \\
      | & \texttt{Forall}(\Pi, \Pi)
          & \text{Dynamical branching: both behaviors are possible} \\
      | & \texttt{Exists}(\Pi, \Pi)
          & \text{Dynamical branching: at least one behavior is possible} \\
      | & \texttt{Assert}(\Phi)
          & \text{Assertion: the formula is true at this point}
  \end{array}$
  \caption{\label{fig:grammar-prog}%
    OCaml grammar for imperative path programs ($\Pi$).
    The symbol $\Phi$ refers to the formulas grammar defined in Figure~\ref{fig:grammar-formula}.}
\end{figure}

Note that in the main OCaml file, the model specification should be written just after the multiplex formula grammar definition,
while the formulas and path programs to process should be defined at the end of the file.

\subsection{Useful Functions}

Taking into consideration the program and the post-condition, the weakest precondition can be computed with the \texttt{wp} function:
$$\texttt{let my\_wp = wp my\_prog my\_post ;;}$$

A simplification can be applied on any formula with the \texttt{simplify} function.
If an initial state and a parametrization are (partially) known, one can also “refine” (that is, strengthen) the weakest precondition with the same function:
$$\texttt{let simpl\_wp = simplify my\_wp known\_vars known\_params ;;}$$
where \texttt{my\_wp} is the weakest precondition (or any other formula to simplify), and \texttt{known\_vars} and \texttt{known\_params} are association lists giving the known values of any number of variables and parameters.
If no such infomation is given, empty lists (\texttt{[]}) are to be provided.
In any case, the \texttt{simplify} function replaces all variables and parameters given in these lists by their provided values, and attempts to perform basic simplifications on the formula, following De Morgan's laws.

At any point, functions are provided to translate a formula (\texttt{string\_of\_formula}), an arithmetic expression (\texttt{string\_of\_expr}) or an imperative path program (\texttt{string\_of\_prog} and \texttt{string\_of\_prog\_indent}) into a pretty-printable string.

Finally, one can translate a formula (typically, the simplified and refined weakest precondition) into Answer Set Programming (ASP) that can be read by Clingo by using function \texttt{write\_example}:
$$\texttt{write\_example my\_wp "file.lp" ;;}$$
This translation is made by creating an ASP atom for each node of the OCaml representation of the formula.
This atom is used in rules such that it reflect its semantics.
For instance, consider the conjunction $f = a \land b$ between some subformulas $a$ and $b$, which ASP representations are atoms \texttt{atom0} and \texttt{atom1}.
This fomrula would be translated into another atom, say \texttt{atom2}, and the conjunction would be encoded by the ASP rule:
$$\texttt{atom2 :- atom0, atom1.}$$
Arithmetic expressions are also translated into their ASP equivalent, while variables and parameters are each assigned to an ASP variable.

Note that the ASP variables that represend parameters are labeled with integers rather than with the explicit names of the resource set $\omega$.
In order to obtain the correspondence between the ASP variable names and the original parameters, one can use the \texttt{asp\_params} function.

See the “Sandbox” part at the end of the main OCaml file for examples on how to use theses functions to obtain results on the model example.

\section{Contents and Usage}

The \hoarefol{} tool is freely available at \url{https://gitlab.cristal.univ-lille.fr/mfolsche/hoare-fol}
under the MIT license\footnote{\url{https://opensource.org/licenses/MIT}}.

The tool can be run in a Unix compatible terminal.
Pease refer to the \texttt{README.txt} file for information on the requirements and how to run the main file.

The \texttt{main.ml} OCaml file contains the main program with somme example applications on the model of Figure~\ref{fig:toy-ocaml}.
It requires OCaml 4.03 to be executed, but the latest version\footnote{OCaml version 4.09 at the time of writing} is recommended.
The command line to run this file is:
$$\texttt{ocaml main.ml}$$

If ASP files are produced by the execution, they can be fed to Clingo\footnote{Produced scripts are intended for Clingo 5. This feature has not been tested with Clingo 4 although the syntax should be compatible. It is compatible with Clingo 3, but it is advised to comment out the \texttt{\#hide.} directive in the produced scripts to hide uninteresting atoms.} with the following command:
$$\texttt{clingo 0 file.lp}$$
Note that the command line option \texttt{0} means “enumerate all solutions”.
It can be replaced by a non-null integer to indicate the maximum number of solutions to enumerate

The provided script \texttt{run-all.sh} allows to run all \texttt{.lp} files with Clingo and store the results in \texttt{.lp.out} files:
$$\texttt{bash run-all.sh}$$

\section{Limitations}

This implementation comes as a proof of concept, and as such still has a number of limitations.

The biggest theoretical limitations are linked to the \texttt{While} loops that are rather difficult to express, and which support is limited in the current version of this tool:
\begin{itemize}
  \item An explicit loop invariant has to be provided for the \texttt{While} loops.
    However, \cite{Bernot2019Genetically} propose a method to automatically infer a weakest invariant with the following approach:
      \begin{itemize}
        \item Start with the most general invariant.
        \item Run the loop and remove values that lead out of boundaries.
        \item Repeat until reaching a fixpoint.
      \end{itemize}
    Since values are finite (variables take bounded discrete values), this is ensured to end.
  \item The weakest preconditions of \texttt{While} loops are expressed as formulas in a special context (\texttt{FreshState}, defining a “fresh” set of variables) which is currently not explored nor simplified buy the \texttt{simplify} function.
    The simplifications should also apply to these formulas, probably with the same simplification rules, but by taking care of not performing refining on variables in such a context.
\end{itemize}
There also are technical limitations regarding the ASP output:
\begin{itemize}
  \item The output of Clingo can be difficult to read, as variables are all encoded with dummy names.
    The \texttt{asp\_params} outputs the correspondence between ASP and model variables, bot does not provide pretty-printing nor replace one with the other in the output.
    A more explicit encoding could be found to ease direct reading of the solutions.
  \item The Clingo solving can be really long for some formulas, especially if there are a lot of solutions.
    This limitation seems hard to fix; working on the formula instead of on the set of solutions seems to be the best alternative in this case.
\end{itemize}
Finally, there also are obvious limitations on the source code itself:
\begin{itemize}
  \item Both model and processings have to be hard-coded in the main file, and at specific locations.
    A parser should be added to load a model from a file, or the main file without examples should be turned into a module.
  \item Functions related to Hoare logic should be purified (they are currently closures on \texttt{vars} and \texttt{mults}, which partly causes the previous limitation).
\end{itemize}

\section{Conclusion}

This paper presents an implementation of \cite{Bernot2019Genetically} which applies Hoare logic to Thomas' formalism in order to infer constraints on the model's parameter values.
It relies on a symbolic representation of logic formulas and imperative programs in order to compute the weakest precondition of a given couple of program and postcondition.
It is written in OCaml and allows an ouput of the formulas in ASP (compatible with Clingo 5) to enumerate solutions.
Although there are theoretical and technical limitations, especially when \texttt{While} loops are involved, or regarding hard-coded features, this work only aims at being the basis of other works that could require such a framework.
This has already been the case with \cite{Behaegel2017Constraint} which re-uses its main concepts, and applies them to a hybrid extension of Thomas' formalism.

\bibliographystyle{apalike}
\bibliography{biblio}
\end{document}